\documentclass[emulateapj,apj]{emulateapj}
\newcommand{\ts}{t_{\rm stop}}
\newcommand{\cs}{c_{\rm s}}
\newcommand{\cd}{c_{\rm d}}
\newcommand{\sd}{\Sigma_{\rm d}}

\shorttitle{Two-Component Secular GI in a Protoplanetary Disk}
\shortauthors{Takahashi, Inutsuka}

\usepackage{natbib, aas_macros}
\begin{document}
\title{Two-Component Secular Gravitational Instability in a
Protoplanetary Disk: 
A Possible Mechanism for Creating Ring-Like
Structures}

\author{Sanemichi Z. Takahashi\altaffilmark{1,2}, Shu-ichiro
Inutsuka\altaffilmark{1}}
\altaffiltext{1}{Department of Physics, Nagoya University, Furo-cho, 
Chikusa-ku, Nagoya, Aichi, 464-8602, Japan;
takahashi.sanemichi@a.mbox.nagoya-u.ac.jp, inutsuka@nagoya-u.jp
}
\altaffiltext{2}{Department of Physics, Kyoto University, Oiwake-cho,
Kitashirakawa, Sakyo-ku, Kyoto 606-8502, Japan;
sanemichi@tap.scphys.kyoto-u.ac.jp}

\begin{abstract}
The instability in protoplanetary disks due to gas-dust 
friction and self-gravity of gas and dust is investigated 
by linear analysis.
In the case where the dust to gas ratio is enhanced and turbulence is week,
the instability grows, even in gravitationally stable disks,
on a timescale of order $10^{4\-- 5}$yr
at a radius of order 100AU.
If we ignore the dynamical feedback from dust grains in the
gas equation of motion,
the instability reduces to the so-called ``secular gravitational instability'',
which was investigated previously as an instability of dust in a fixed
 background gas flow.
In this work, we solve the equations of 
motion for both gas and dust consistently and
 find that long-wavelength perturbations are stable, in contrast to the
 secular gravitational instability in the simplified treatment.
This may indicate that we should not neglect small terms in equation of
 motion if the growth rate is small.
The instability is expected to form ring structures 
in protoplanetary disks.
The width of the ring formed at a radius of 100 AU is a few tens of AU.
Therefore, the instability is a candidate for the formation mechanism 
of observed ring-like structures in disks.
Another aspect of the instability is the accumulation of dust
 grains, and hence the instability may play an important role 
in the formation of planetesimals, rocky protoplanets, and cores of gas
 giants located at radii $\sim$100 AU.
If these objects survive the dispersal of the gaseous component of the disk,
they may be the 
origin of debris disks.
\end{abstract}


\def\bm#1{\mbox{\boldmath $#1$}}

\section{INTRODUCTION}
Since planets are expected to form in protoplanetary disks, 
the formation and evolution of disks affect the 
planet formation process.
In particular, terrestrial planets and cores of gas giants are expected to
be formed via dust growth \cite[]{1985prpl.conf.1100H}.
Therefore, the dynamics of the dust in protoplanetary disks
is essential to planet formation.

Recently, high-angular-resolution direct imaging of protoplanetary disks 
has become available.
The observations reveal that ring structures are formed in
protoplanetary disks
\cite[e.g.][]{2006ApJ...636L.153F,2013arXiv1309.7400F,2007A&A...469L..35G,
2010ApJ...725.1735I,
2012ApJ...747..136I,
2013ApJ...775...30I,
2011ApJ...732...42A,2011ApJ...729L..17H,2012ApJ...758L..19H,
2012ApJ...753...59M,2012ApJ...760L..26M,
2013Natur.493..191C,2013Sci...340.1199V}.
The formation mechanism of such structures remains unknown.
One candidate for the formation mechanism is the 
density gap formed by gravitational
interaction between the disk and unseen giant planets. 
\cite[e.g.][]{1986ApJ...309..846L,
1993prpl.conf..749L,1996ApJ...460..832T,
2012ARA&A..50..211K}.
However, planets have not been observed in most of the gaps.
Moreover, \cite{2011ApJ...729...47Z} have mentioned that 
several planets are needed to make the observed wide gap.
If planets are present in the gap, their maximum mass is
estimated to be about five times the Jupiter mass \cite[]{2011ApJ...729L..17H,
2012ApJ...758L..19H,
2013Natur.493..191C}.
Since the sensitivity of observations will increase, the 
maximum mass of the unseen planets will decrease and the 
observations will make clear whether
rings are formed by planets.

In this work, we investigate a ring-formation mechanism without
planets.
We focus on instability due to gas-dust friction.
This has been well studied, 
especially in the context of planetesimal formation.
The streaming instability \cite[]{2005ApJ...620..459Y,
2007ApJ...662..613Y,2007ApJ...662..627J}
occurs via dust motion towards a central star (radial drift).
The secular gravitational instability \cite[secular GI,][]{2000orem.book...75W,
2011ApJ...731...99Y,2012ApJ...746...35M}
is the gravitational collapse of dust due to gas-dust friction.
In a dust-rich disk
the dust has large effect on the disk.
Instability in dust rich disks has been investigated by 
\cite{1981A&A....98..173C}.
Recently, \cite{2012arXiv1204.6322L,
2013Natur.499..184L} investigated the instability that occurs 
when gas is heated by dust via photoelectric heating in the debris disk,
which contains more dust than gas.

In this work, we use full two-fluid equations for gas and dust 
in a typical gas-rich disk.
We perform a local linear stability analysis and  discuss
whether the instability can explain the formation of the
ring structures observed in protoplanetary disks.
Although secular GI may also form dust rings,
the analysis of secular GI only 
uses equations only for dust so
secular GI cannot explain the observed ring structures of gas and dust
\cite[]{2012ApJ...753...59M,
2013Natur.493..191C,
2013arXiv1309.7400F,2013Sci...340.1199V}.
Therefore, a linear analysis using equations for both gas and dust is 
needed to explain protoplanetary disk ring structures.
We call the instability discussed in this paper ``two-component
secular GI''.
In this work, we also take into account the effect of gas self-gravity.

This paper is organized as follows.
The basic equations for gas and dust for the linear analysis are given
in  Section \ref{basic_equations}.
In Section \ref{results} we derive the dispersion relation and 
parameter dependence of the maximum growth rate.
We discuss the condition for the instability to grow in protoplanetary
disks in Section \ref{discussion}.
A summary is given in Section \ref{conclusion}.

\section{BASIC EQUATIONS}
\label{basic_equations}
We investigate instability due to gas-dust friction in
protoplanetary disks by using two fluid equations for both 
gas and dust.
We focus on purely horizontal motions. 
We use the equations of continuity and motion for both gas and dust, and
thePoisson equation:
\begin{equation}
 \frac{\partial \Sigma}{\partial t} + \bm{\nabla}\cdot(\Sigma
  \bm{u})=0,
\end{equation}
\begin{eqnarray}
\Sigma\left(\frac{\partial \bm{u}}{\partial
       t}+(\bm{u}\cdot\bm{\nabla})\bm{u}\right)&=&-c_s^2\bm{\nabla}\Sigma-\Sigma\bm{\nabla}\left(\Phi
- \frac{GM_*}{r}\right) \nonumber \\
&&+ \frac{\Sigma_d(\bm{v}-\bm{u})}{t_{\rm stop}},
\label{eq:eom_gas}
\end{eqnarray}
\begin{equation}
  \frac{\partial \Sigma_d}{\partial t} + \bm{\nabla}\cdot(\Sigma_d
   \bm{v})=D\nabla ^2\Sigma,
\end{equation}
\begin{eqnarray}
\sd\left(\frac{\partial \bm{v}}{\partial
       t}+(\bm{v}\cdot\bm{\nabla})\bm{v}\right)=-\Sigma_d\bm{\nabla}\left(\Phi
- \frac{GM_*}{r}\right) \nonumber \\
+\frac{\Sigma_d(\bm{u}-\bm{v})}{t_{\rm stop}},
\label{eq:eom_dust}
\end{eqnarray}
\begin{equation}
 \nabla^2\Phi=4\pi G(\Sigma+\Sigma_d)\delta(z),
\end{equation}
where $\Sigma$ and $ {\bm u}$ are surface density and velocity of gas, 
$\sd$ and ${\bm v}$ are surface density and velocity of dust,
$\cs$ is the sound speed of gas, $D$ is the diffusivity of the dust
due to the gas turbulence,
$M_*$ is the central star mass, and 
$\ts$ is the stopping time of a dust particle.

We adopt a local shearing box model
\cite[e.g.][]{1965MNRAS.130..125G,1987MNRAS.228....1N}.
We focus on the neighborhood of a point 
$(r,\theta )=(r_0, \Omega t)$ in cylindrical coordinates.
The local radial and azimuthal coordinates are
$(x,y)=(r-r_0,r_0(\theta -\Omega t))$.
We adopt the Keplerian frequency, $\Omega$,  at $r=r_0$
; $\Omega =\sqrt{GM_*/r_0^3}$.
For simplicity, we assume axisymmetry.

We assume a steady state background 
 with uniform surface density;
$\Sigma_0 , \Sigma_{{\rm d}0} = {\rm const}$;
dust-to-gas mass ratio $\epsilon=\Sigma_{{\rm d}0}/\Sigma_0$;
and Keplerian rotation
$u_{x0}=v_{x0}=0,u_{y0}=v_{y0}=(-3/2)\Omega_0 x$.
We decompose the physical quantities into background values
and small
perturbations proportional to $\exp[ikx-i\omega t]$.
The linearized equations are given as follows:
\begin{equation}
 -i\omega \delta \Sigma + ik \Sigma_0 \delta u_x=0,
\label{eq:eoc_gas_l}
\end{equation}
\begin{equation}
-i\omega  \delta u_x -2\Omega \delta u_y =
 -\cs^2\frac{ik\delta\Sigma}{\Sigma_0 }
 - ik \delta\Phi + \frac{\epsilon (\delta v_x- \delta u_x)}{t_{\rm stop}},
\label{eq:eom_gas_x_l}
\end{equation}
\begin{equation}
-i\omega \delta u_y  + \frac{\Omega}{2} \delta u_x=
\frac{\epsilon (\delta v_y-\delta u_y)}{t_{\rm stop}},
\label{eq:eom_gas_y_l}
\end{equation}
\begin{equation}
  -i\omega \delta \sd + ik \epsilon \Sigma_0  \delta v_x=-Dk^2 \delta \sd,
\label{eq:eoc_dust_l}
\end{equation}
\begin{equation}
-i\omega \delta v_x -2\Omega \delta v_y =
-ik\delta \Phi+ \frac{\delta u_x-\delta v_x}
{t_{\rm stop}},
\label{eq:eom_dust_x_l}
\end{equation}
\begin{equation}
-i\omega \delta v_y +\frac{\Omega}{2}  \delta v_x =
\frac{\delta u_y-\delta v_y}{t_{\rm stop}},
\label{eq:eom_dust_y_l}
\end{equation}
\begin{equation}
 \delta \Phi=- \frac{2\pi G (\delta \Sigma+\delta \sd )}{|k|}.
\label{eq:poisson_l}
\end{equation}
Hereafter we use the growth rate of the instability
$n\equiv -i\omega$ instead of the frequency $\omega$.

\section{RESULTS}
\label{results}
From the linearized equations derived in Section \ref{basic_equations},
we derive the dispersion relation for the instability. 
The dispersion relation given by Equations (\ref{eq:eoc_gas_l}) to
(\ref{eq:poisson_l}) is sixth order.
Therefore, the dispersion relation implies six possible modes.
Four modes are decaying oscillations with ${\rm Re}[n]<0$ and 
${\rm Im}[n]\neq 0$.
The other two modes can be unstable.
Hereafter we investigate the unstable modes further.
\subsection{Apploximate Dispersion Relation}
\begin{figure}
 \epsscale{1}
\plotone{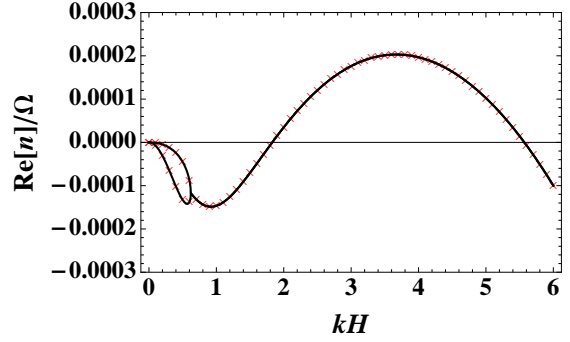}
\caption{Dispersion relation of the unstable mode for
$\ts\Omega=0.01$, $Q=3,\ \epsilon=0.1,\ D=10^{-4}\cs^2\Omega^{-1}$.
The horizontal axis is the normalized wavenumber, $kH$, where
$H \equiv \cs/\Omega$ is the scale height of the gas disk. 
The vertical axis is the normalized growth rate of the instability, 
${\rm Re}[n]/\Omega$.
The crosses show the exact solution given by Equations 
(\ref{eq:eoc_gas_l}) to (\ref{eq:poisson_l}).
The Solid line shows the approximate solution given by Equation
(\ref{eq:desp_2fluid_SGI}).
}
\label{fig:t01Q3e01cd01sol2}
\end{figure}

We provide approximate solutions of the dispersion relation
and the condition for the instability.
In the case $n \ll \ts^{-1}, \Omega$ and $\ts \Omega \ll 1$ are satisfied,
the dispersion relation is approximately given by the following
quadratic equation,
\begin{eqnarray}
&&\left(\frac{1+\epsilon}{\ts}\right)^2\left[\Omega ^2 -2\pi G(1+\epsilon)
 \Sigma _0 k +\frac{\cs^2 k^2}{1+\epsilon}\right]n^2\nonumber \\
&& +\left\{ \frac{\epsilon \cs ^2 k^2}{\ts}\left[\Omega^2 -2\pi G
		    (1+\epsilon)\Sigma k\right] \nonumber \right.\\
&&\left.+Dk^2\left(\frac{1+\epsilon}{\ts}\right)^2\left[\Omega^2 -2\pi G
      \Sigma_0 k +\frac{\cs^2k^2}{1+\epsilon}\right]
    \right\}n\nonumber \\
&&+\frac{\epsilon \cs^2 k^2}{\ts}\Omega ^2 Dk^2=0.
\label{eq:desp_2fluid_SGI}
\end{eqnarray}
Fig \ref{fig:t01Q3e01cd01sol2} shows the real part of 
the exact dispersion relation of the
unstable modes for $\ts\Omega=0.01,\ \epsilon=0.1,\
D=10^{-4}\cs^2\Omega^{-1}, Q=3$ and 
the real part of 
the approximate solution given by Equation (\ref{eq:desp_2fluid_SGI}),
 where $Q\equiv \cs\Omega/\pi G \Sigma$ is 
Toomre's parameter.
The figure indicates good agreement between the exact dispersion relation
and the approximate one.

\subsection{Conditions for Instability}
There are two cases where the solutions of the Equation
(\ref{eq:desp_2fluid_SGI}) have at least one positive real part.
One condition is given by
\begin{equation}
\Omega^2-2\pi G(1+\epsilon) \Sigma _0 k +\frac{\cs^2k^2}{1+\epsilon}<0.
\label{eq:condition_for_unstable1}
\end{equation}
The other case is given by
\begin{equation}
\Omega^2-2\pi G(1+\epsilon) \Sigma _0 k +\frac{\cs^2k^2}{1+\epsilon}>0,
\label{eq:condition_for_unstable2-1}
\end{equation}
and
\begin{eqnarray}
&& \frac{\epsilon \cs^2 k^2}{\ts}[\Omega ^2 -2 \pi
  G(1+\epsilon)\Sigma_0 k]\nonumber\\
&&+Dk^2\left(\frac{1+\epsilon}{\ts}\right)^2
  \left[\Omega^2 - 2 \pi G\Sigma_0 k +\frac{\cs^2
   k^2}{1+\epsilon}\right]<0.
\label{eq:condition_for_unstable2-2}
\end{eqnarray}
Here we use that the last term in the left hand side of 
the Equation(\ref{eq:desp_2fluid_SGI}) is always positive.
From Equation (\ref{eq:condition_for_unstable1}), 
we obtain the condition for instability
\begin{equation}
 Q=\frac{\cs \Omega}{\pi G\Sigma} < (1+\epsilon)^{3/2}.
\label{eq:condition_for_modified_GI}
\end{equation}
This condition is similar to the condition for the gravitational 
instability of the gas disk and it is completely same in the case 
$\epsilon =0$. 
This instability is the gravitational instability of the gas
and dust disk.
In the case Equation (\ref{eq:condition_for_modified_GI}) is not 
satisfied, the disk is unstable only if
Equation (\ref{eq:condition_for_unstable2-2}) is satisfied.
From Equation (\ref{eq:condition_for_unstable2-2}), the condition for
instability is given by 
\begin{equation}
 \frac{\cs ^2 \Omega ^2 D\left[\frac{\epsilon}{1+\epsilon}\ts \cs ^2 +
			   D(1+\epsilon)\right]}
{(\pi G \Sigma_0)^2[\epsilon \ts \cs^2 + D(1+\epsilon)]^2}<1.
\label{eq:condition_for_insrabiliy2}
\end{equation}

In the case $\epsilon \ll 1$ and $\epsilon \ts \cs^2 \ll D$,
Equation (\ref{eq:condition_for_insrabiliy2}) is rewritten as
\begin{equation}
 Q\lesssim(1+\epsilon)^{1/2}.
\end{equation}
This condition is not satisfied in the case Equation
(\ref{eq:condition_for_unstable1}) is not satisfied.
Therefore, the disk is stable when $Q>(1+\epsilon)^{3/2}$ and 
$\epsilon \ts \cs^2 \ll D$ is satisfied.

In the case $\epsilon \ll 1$ and $\epsilon \ts \cs^2 \gg D$,
Equation (\ref{eq:condition_for_insrabiliy2}) is rewritten as
\begin{equation}
  \frac{1}{\epsilon (1+\epsilon)\ts}\frac{D \Omega^2}{(\pi G \Sigma_0
  )^2}\lesssim 1,
\end{equation}
or 
\begin{equation}
  \frac{{\bar D}Q^2}{\epsilon (1+\epsilon)\ts\Omega}\lesssim 1,
   \label{eq:app_condition_for_instabiity2}
\end{equation}
where ${\bar D} \equiv D \cs^{-2} \Omega$ is a normalized diffusivity.

 If we ignore  dynamical feedback from dust grains in the gas equation
 of motion,
 the present instability reduces to the so-called  secular GI
\cite[e.g.,][]{2000orem.book...75W}.
In that case long-wavelength perturbations are always unstable 
to the secular GI,
independent of the parameters $D,\ \Sigma_{\rm d0}$, and $\Omega$.
However, the
feedback of the friction force from dust to gas 
makes the long-wavelength perturbations stable,
even in disks with a low dust-to-gas mass ratio.
Moreover, the perturbations are stable 
for all wavelengths
if Equation (\ref{eq:app_condition_for_instabiity2})
is not satisfied.
In this way the present instability is different from the secular GI.

\section{DISCUSSION}
\label{discussion}
\subsection{Turbulent Viscosity, Dust Velocity Dispersion and 
Disk Thickness}
\label{other_effects}
In Sections \ref{basic_equations} and \ref{results}, we neglected
turbulent viscosity, dust velocity dispersion, and disk thickness,
 for simplicity.
We take into account those effects in this section.

In the turbulent disk, the momentum of the gas is also diffused by
the turbulent viscosity.
Coefficient of kinematic viscosity is given by $\nu = \alpha \cs^2
/\Omega$, where $\alpha$ is a dimensionless measure of turbulent
intensity \cite[]{1973A&A....24..337S}.
Taking into account the turbulent viscosity,
we can rewrite Equations (\ref{eq:eom_gas}) as follows:
\begin{eqnarray}
&\Sigma&\left(\frac{\partial u_i}{\partial
       t}+u_k \frac{\partial u_i}{\partial x_k}\right) \nonumber \\
&=&-c_s^2\frac{\partial \Sigma}{\partial x_i}-\Sigma\frac{\partial
}{\partial x_i}\left(\Phi
- \frac{GM_*}{r}\right)
+ \frac{\Sigma_d(v_i-u_i)}{t_{\rm stop}}\nonumber \\
&&+\frac{\partial}{\partial x_k}
\left[\Sigma \nu \left(\frac{\partial u_i}{\partial x_k}
                       +\frac{\partial u_k}{\partial x_i}
		       -\frac{2}{3}\delta_{ik}\frac{\partial
		       u_l}{\partial x_l}\right)\right].
\end{eqnarray}
Then linearized equations of this equation are given as follows:
\begin{eqnarray}
-i\omega  \delta u_x -2\Omega \delta u_y &=&
 -\cs^2\frac{ik\delta\Sigma}{\Sigma_0 }
 - ik \delta\Phi + \frac{\epsilon (\delta v_x- \delta u_x)}{t_{\rm
 -stop}} \nonumber\\
 &&-\left(\xi +\frac{4}{3}\nu\right)k^2\delta u_x,
\label{eq:eom_gas_x_l_vis}
\end{eqnarray}
\begin{equation}
-i\omega \delta u_y  + \frac{\Omega}{2} \delta u_x=
\frac{\epsilon (\delta v_y-\delta u_y)}{t_{\rm stop}}
-\nu k^2 \delta u_y -ik\frac{3\nu\Omega}{2\Sigma_0}\delta \Sigma.
\label{eq:eom_gas_y_l_vis}
\end{equation}
In this case, 
viscous overstability also appears \cite[]{1995Icar..115..304S},
but in the case $\alpha$ is sufficiently small, 
the growth rate of the viscous overstability 
is much smaller than that
of two-component secular GI. 
In this paper we neglect viscous overstability and focus on 
the two-component secular GI.
The relation between turbulent viscosity and dust diffusivity is 
given by \cite{2007Icar..192..588Y} \cite[see also][]{2012ApJ...746...35M};
\begin{equation}
 {\bar D} = \frac{1+\ts\Omega +4(\ts\Omega)^2}{[1+(\ts\Omega)^2]^2}\alpha.
\end{equation}

In a turbulent disk, velocity dispersion of the dust is estimated by 
\cite{2007Icar..192..588Y};
\begin{equation}
 \cd^2 =  \frac{1+2\ts\Omega +(5/4)(\ts\Omega)^2}{[1+(\ts\Omega)^2]^2}
\alpha\cs^2.
\end{equation}
The velocity dispersion of the dust appears in the equations of motion
for the dust as pressure-like terms.
Taking into account the velocity dispersion,
we can rewrite Equation (\ref{eq:eom_dust}) as follows:
\begin{eqnarray}
\sd\left(\frac{\partial \bm{v}}{\partial
       t}+(\bm{v}\cdot\bm{\nabla})\bm{v}\right)&=&-\cd^2\nabla\sd
-\Sigma_d\bm{\nabla}\left(\Phi
- \frac{GM_*}{r}\right) \nonumber \\
&&+\frac{\Sigma_d(\bm{u}-\bm{v})}{t_{\rm stop}}.
\end{eqnarray}
Then the linearized equation of this equation is given as follows:
\begin{equation}
-i\omega \delta v_x -2\Omega \delta v_y =
-\cd^2 \frac{ik \delta \sd}{\Sigma_0} -ik\delta \Phi
+ \frac{\delta u_x-\delta v_x}
{t_{\rm stop}}.
\label{eq:eom_dust_x_l_p}
\end{equation}

Equation (\ref{eq:poisson_l}) gives the gravitational potential
perturbation in the infinitesimally thin disk.
However, the most unstable wavelength of the two-component secular GI is 
of order $H^{-1}$ and infinitesimally thin approximation is not good 
approximation.
The gravitational potential perturvation in the case $k\lesssim H^{-1}$ is
approximately given by \cite{1970ApJ...161...87V} and
\cite{1984prin.conf..513S}; 
\begin{equation}
 \delta \Phi = - 2\pi G\left(\frac{\delta \Sigma}{1+kH}+
\frac{\delta \sd}{1+ k H_d}\right),
 \label{eq:poisson_l_disk_hight}
\end{equation}
where $H_d \sim \sqrt{\alpha/\ts\Omega} H$ is the dust scale height 
\cite[]{1993Icar..106..102C}.

\begin{figure}
 \epsscale{1}
 \plotone{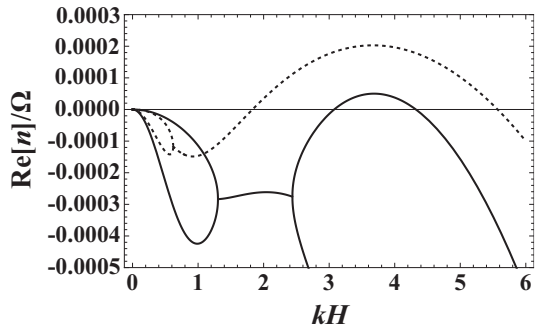}
 \caption
{Dispersion relations of the unstable modes for 
$\ts\Omega=0.01$, $Q=3,\ \epsilon=0.1,\ D=10^{-4}\cs^2\Omega^{-1}$.
Solid lien shows the dispersion relation given by Equations 
(\ref{eq:eoc_gas_l}), (\ref{eq:eoc_dust_l}), (\ref{eq:eom_dust_y_l}), 
(\ref{eq:eom_gas_x_l_vis}), (\ref{eq:eom_gas_y_l_vis}), 
(\ref{eq:eom_dust_x_l_p}), and (\ref{eq:poisson_l_disk_hight}).
Dashed line shows dispersion relation shown in Figure
\ref{fig:t01Q3e01cd01sol2}.
The growth rate of the instability is decreased 
by the  turbulent viscosity, dust velocity dispersion, and disk
 thickness.
}
 \label{fig:comp}
\end{figure}

Using Equations (\ref{eq:eoc_gas_l}), (\ref{eq:eoc_dust_l}), 
(\ref{eq:eom_dust_y_l}), 
(\ref{eq:eom_gas_x_l_vis}), (\ref{eq:eom_gas_y_l_vis}), 
(\ref{eq:eom_dust_x_l_p}), and (\ref{eq:poisson_l_disk_hight}),
we perform the linear stability analysis again.
Figure \ref{fig:comp} shows the dispersion relation given by Equations
(\ref{eq:eoc_gas_l}), (\ref{eq:eoc_dust_l}),  
(\ref{eq:eom_dust_y_l}), 
(\ref{eq:eom_gas_x_l_vis}), (\ref{eq:eom_gas_y_l_vis}), 
(\ref{eq:eom_dust_x_l_p}), and (\ref{eq:poisson_l_disk_hight}), and
the dispersion relation shown in Figure \ref{fig:t01Q3e01cd01sol2}.
The growth rate of the instability is decreased 
due to turbulent viscosity, dust velocity dispersion, and disk thickness.

\begin{figure}
 \epsscale{1}
 \plotone{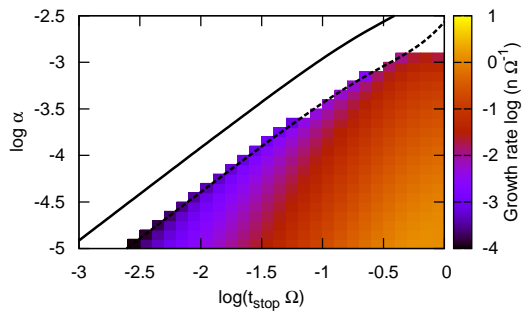}
 \caption{
 Maximum growth rate of the unstable modes for $Q=3$ and $\epsilon
 =0.1$. The horizontal axis is 
normalized stopping time and the vertical axis is turbulence parameter, 
$\alpha$.
The white region shows negative growth rate. 
The solid line shows the condition for instability given by Equation 
(\ref{eq:app_condition_for_instabiity2}) and
the dashed line is given by Equation
 (\ref{eq:app_condition_for_instabiity3}). 
}
\label{fig:maximum_growth_rate}
\end{figure}

Figure \ref{fig:maximum_growth_rate} shows the maximum growth rate of the unstable modes.
Solid line shows the condition for instability given by Equation 
(\ref{eq:app_condition_for_instabiity2}) and
dashed line is given by 
\begin{equation}
   \frac{3{\bar D}Q^2}{\epsilon (1+\epsilon)\ts\Omega} = 1.
   \label{eq:app_condition_for_instabiity3}
\end{equation}

The maximum $\alpha$ for the instability is well fitted by the Equation  
 (\ref{eq:app_condition_for_instabiity3}). 
Due to the effect of the turbulent viscosity, velocity dispersion of the
dust and disk thickness, the maximum $\alpha$ for the instability 
is decreased by a factor of three.

\subsection{Dust Growth and Radial Drift}
The growth rate of the instability depends on the stopping time, $\ts$.
The stopping time is given as a function of the dust
radius, $a$, and gas density, $\rho_{\rm g}$:
\begin{equation}
\ts = 
\left\{
\begin{array}{ll}
 \frac{\rho_{\rm int}a}{\rho_{\rm g} \cs} & \frac{a}{l_{\rm g}}<\frac{3}{2} \\
 \frac{2\rho_{\rm int}a^2}{3\rho_{\rm g}\cs l_{\rm g} } &
  \frac{a}{l_{\rm g}}>\frac{3}{2} \\
\end{array} 
\right.
\label{eq:st}
\end{equation}
where $\rho_{\rm int}$ is the internal density of dust and 
$l_{\rm g}$ is the mean free path of the gas.
Therefore, the stopping time increases as the radius of dust becomes
large due to coagulation.

The growth timescale of the dust is given by
\begin{eqnarray}
  t_{\rm grow}&=&\left(\frac{d \ln m}{dt}\right)^{-1}\nonumber\\
&=&\frac{4}{3}\frac{\rho_{\rm int}a}{\Delta v \rho_{\rm d}}.
\label{eq:tgrow}
\end{eqnarray}
where $\rho _{\rm d}$ is the dust density and $\Delta v$ is the relative
velocity of the dust.
In the case $a<3l_{\rm g}/2$, from Equations (\ref{eq:st}) and
 (\ref{eq:tgrow}), the growth timescale is given by
\begin{equation}
 t_{\rm grow} =\frac{4}{3}\frac{\rho_{\rm g}\cs \ts}{\Delta v \rho_{\rm
  d}}.
\end{equation}
Using the relations 
$\Delta v = \sqrt{\ts\Omega}\ \cd
= \sqrt{\alpha\ts\Omega}\ \cs$ \cite[e.g.][]{2010AREPS..38..493C} 
and
$\rho_{\rm g}/\rho _{\rm d} = (\Sigma/\sd)(H_{\rm d}/H)$,
we obtain the growth timescale as follows \cite[]{2005ApJ...623..482T};
\begin{equation}
 t_{\rm grow} = \frac{4}{3\epsilon \Omega}.
\end{equation}
In Section \ref{basic_equations}, we adopted a uniform surface density
and Keplerian rotation for both gas and dust as the background state.
However, the surface density of the gas and dust depend on the radius of
the disk, for example $\Sigma \propto r^{-3/2}$ for the minimum mass Solar
Nebula \cite[]{1981IAUS...93..113H}.
The gas is supported by pressure and rotates slightly slower than 
the Kepler velocity: $\Omega_{\rm gas}=(1-\eta)\Omega$, $\eta >0$.
Therefore, dust loses angular momentum due to gas-dust friction
and drifts towards the central star.
The radial drift speed is given by \cite[]{1986Icar...67..375N}
\begin{equation}
 v_r=-2\eta r\Omega\frac{1}{1+\epsilon}\frac{\ts\Omega}{1+(\ts\Omega)^2}.
\end{equation}
In the case of the minimum mass Solar Nebula, 
$\Sigma\propto r^{-3/2}$ and $T = 280 (r/1{\rm AU})^{-1/2}$K, 
we obtain 
\begin{eqnarray}
 \eta &=& \frac{1}{2}\left(\frac{\cs}{r\Omega}\right)^2\frac{d\ln P}{d \ln
  r}\nonumber\\
&\simeq& 1.8\times10^{-3}\left(\frac{r}{1 {\rm AU}}\right)^\frac{1}{2}.
\end{eqnarray}
Therefore, at 100 AU we obtain $\eta\sim 10^{-2}$.
The radial drift timescale is given by
\begin{equation}
 t_{\rm dri}=
  \frac{r}{|v_r|}=\frac{1+\epsilon}{2\eta\Omega}
\frac{1+(
  \ts\Omega)^2}{\ts\Omega}.
\end{equation}

We assume that dusts drift inward without coagulating when $t_{\rm
grow} \lesssim t_{\rm dri}/30$ \cite[]{2012ApJ...752..106O}.
Then we obtain the critical stopping time;
\begin{equation}
 t_{\rm stop,crit} \simeq 0.13 \Omega^{-1}
\left(\frac{\epsilon}{0.1}\right)
\left(\frac{\eta}{0.01}\right)^{-1}
\end{equation}
where we assume $\epsilon \ll 1$ and $\ts\Omega \ll 1$.
When the dust is small and $\ts$ is smaller than $t_{\rm stop,crit}$, 
then $t_{\rm grow}$ is smaller than $t_{\rm dri}/30$.
In this case dust grows quickly and radial drift is inefficient.
The stopping time becomes larger as the dust becomes larger, and 
dust drifts inward when $t_{\rm grow} = t_{\rm dri}/30$ is
satisfied.
Since the gas density of the disk is larger at smaller disk radius,
the stopping time of dust coming from outer disk 
radii is smaller than $t_{\rm stop,crit}$ (Equation(\ref{eq:st})). 
The dust grows again to satisfy $\ts =t_{\rm stop,crit}$ and it 
drifts further inwards.
As a result, dust satisfying $\ts=t_{\rm stop,crit}$ exists
at each radius.
Equation (\ref{eq:app_condition_for_instabiity3}) is rewritten
by using $\ts = t_{\rm stop,crit}$;
\begin{equation}
 \left(\frac{\alpha}{4\times 10^{-5}}\right)
  \left(\frac{\epsilon}{0.1}\right)^{-2}
\left(\frac{Q}{10}\right)^{2}
\left(\frac{\eta}{0.01}\right)
\lesssim
1.
\end{equation}
It is quite difficult to satisfy this condition in the MRI turbulent
disk ($\alpha \sim 0.01$).
The maximum $\alpha$ for the instability given by this is smaller than the
$\alpha_{\rm max}$ given by \cite{2011ApJ...731...99Y}.
Therefore, the effect of the back reaction from the dust to gas 
makes it difficult to form the planetesimal due to the secular
GI even in a region without MRI.

\subsection{Ring Structure Formation in Protoplanetary Disks}
Observations of protoplanetary disks show that 
ring structures form at a radius of about 100 AU.
We evaluate the most unstable wavelength and the growth timescale 
of the instability at a radius of 100 AU for the case of 
a $1 M_{\rm \odot}$ central star,
a temperature of 28K, and a dust stopping time 
$\ts=t_{\rm stop,crit} = 1.3\times 10^{-1}\Omega^{-1}$.
We assume the case $Q=3$, corresponding to a marginally 
gravitationally stable disk, 
a high dust-to-gas mass ratio, $\epsilon =0.1$,
and $\alpha = 4 \times 10^{-4}$.
These parameters corresponding to $\eta = 10^{-2}$.
Then, the most unstable wavelength is about  13 AU,
 and the growth timescale is about $2 \times 10^4$ yr.
These results are consistent with the observation because the
disk lifetime is about $10^6$ yr and the observed ring width is a few
tens of AU.
The dust radius that corresponds to a stopping time $\ts
 = 0.13\Omega ^{-1}$ in this disk is about 4 mm. 

The amplitude of the dust surface density eigenfunction 
is larger than that of gas surface density.
This means that the dust-to-gas mass ratio increases
as the perturbation grows.
Therefore, the instability forms a ring-like structure
where the dust is concentrated.
If the dust is sufficiently concentrated, gravitational instability 
of the ring will occur and 
the resulting gravitational collapse and fragmentation of the ring 
will form planetesimals.
Therefore, the present
 instability is important for the formation of planetesimals and rocky planet
at outer disk radii of around 100AU.
Planetesimals or rocky planets located at outer radii
may survive disk gas dispersal.
They occasionally collide with each other over a long timescale 
and provide small dust
if the collisions are destructive.
After gas dispersal, small dust is replenished by the collisional 
destruction of planetesimals, and rocky planets 
are expected to form a debris disk of solid particles. Hence, 
these objects may provide a unique origin for debris disks,
because the dust grains in debris disks have a short lifetime and the
debris disk requires continuous replenishment of dust grains. 
Therefore, the two-component secular GI may play an important role in the
 formation of debris disks.

Some of the observed rings are not axisymmetric.
To investigate non-axisymmetric accumulation of gas and dust in
the azimuthal direction,
we need a linear analysis for non-axisymmetric modes
Observations of the dust continuum show the dust surface density
is about 10 times larger than that of the gap.
To compare the resultant structure of the instability with 
observations, we have to investigate the non-linear effects of 
the instability.
In this work we treat the dust as a fluid that diffuses out by the gas
turbulence with diffusivity $D$.
Since the condition for the instability depends on $D$, a more 
realistic treatment of the dust is important for further analysis.

If ring structures formed by the two-component secular GI
are observed, it may be an indicator of the dust concentration and 
week turbulence in the disk.

\subsection{Comparison with the Previous Works}
The secular GI has been investigated for
 planetesimal formation.
\cite{1976fras.conf....1W, 2000orem.book...75W}
has performed the linear stability analysis of the the secular
GI taking into account the velocity dispersion
 of the dust particles (see Section \ref{other_effects}), but not
 taking into account the diffusion due to the turbulence.
 \cite{2011ApJ...731...99Y} takes into account the diffusion due to the
 turbulence and found that the dust diffusion decreases the growth rate
 of the secular GI.
These previous works solved only the equations for dust in the Keplerian
 rotation gas disk. They did not taking into account the back reaction
 from the dust to the gas since the surface density of the dust is usually much
 smaller than that of gas.
In this work, we take into account the back reaction and solve equations
 for both gas and dust. 
We have found that 
 when the growth timescale of the
 instability is much longer than the dynamical timescale,
the back reaction is not
 negligible even if the surface density of the dust is
 smaller than that of gas.
In such a case, Coriolis force stabilizes the long wavelength
 perturbations.
The condition for the two-component secular GI is approximately
 given by Equation (\ref{eq:condition_for_unstable2-2}).
Therefore, the terms proportional to $\Omega^2$ stabilize the long
wavelength perturbations.
In the following, we show that these terms come from the Coriolis force.
From the equations of motion for gas and dust, the equations for the relative
velocity are given by
\begin{equation}
 -2\Omega(\delta v_y - \delta u_y)
= \cs^2 \frac{i k \delta
  \Sigma}{\Sigma_0}
+\frac{1+\epsilon}{\ts}(\delta u_x - \delta v _x),
\label{eq:rev_x}
\end{equation}
\begin{equation}
 \frac{\Omega}{2}(\delta v_x - \delta
  u_x) = \frac{1+\epsilon }{\ts}(\delta u_y - \delta v_y),
\label{eq:rev_y}
\end{equation}
where we use $|\omega| \ll \ts^{-1}$ (terminal velocity
approximation).
Using (\ref{eq:eoc_gas_l}) and (\ref{eq:rev_y}) to eliminate $\delta
\Sigma$, $\delta u_y$ and $\delta v_y$ from (\ref{eq:rev_x}),
we obtain 
\begin{eqnarray}
&&\left[\Omega^2 + \left(\frac{1+\epsilon}{\ts}\right)^2\right]\delta v_x\nonumber\\
&&=\left[\Omega^2 +
  i\frac{\cs^2k^2}{\omega}\frac{1+\epsilon}{\ts}+\left(\frac{1+\epsilon}{\ts}\right)^2\right]\delta
u_x.
\label{eq:rev_x-2}
\end{eqnarray}
In the case that the dust is small and hence, $\Omega \ll \ts^{-1}$,
Equation (\ref{eq:rev_x-2}) is rewritten as
\begin{equation}
\delta v_x
=\left(\frac{\cs^2k^2}{n}\frac{\ts}{1+\epsilon}+1\right)\delta u_x.
\end{equation}
Hereafter, we assume that pressure term $\cs^2k^2$ is large and 
$\cs^2k^2\ts/(n(1+\epsilon))\gg 1$ is satisfied.
This condition is satisfied for example in the case $Q=3$ and $kH \sim 2$.
Using this approximation, we obtain
\begin{equation}
\delta v_x
 =\frac{\cs^2k^2}{n}\frac{\ts}{1+\epsilon}\delta u_x 
\gg \delta u_x.
\label{eq:dux_dvx}
\end{equation}
This means that the radial motion of the gas is suppressed by the pressure
and it is negligible compared to the radial motion of the dust.
Therefore, the relative velocity in the y-direction is given by
\begin{equation}
 \delta u_y - \delta v_y \simeq \frac{\Omega\ts}{2(1+\epsilon)}\delta
  v_x \ll \delta v_x
\label{eq:duy_dvy}
\end{equation}
From Equations (\ref{eq:eom_gas_y_l}) and (\ref{eq:eom_dust_y_l}),
we also obtain the
equation of the centroid velocity in the y-direction
\begin{equation}
 n( \delta u_y + \epsilon \delta v_y ) + \frac{\Omega}{2}(\delta u_x +
  \epsilon \delta v_x) =0.
\label{eq:cev_y}
\end{equation}
Using (\ref{eq:dux_dvx}) and (\ref{eq:duy_dvy}), we can rewrite Equation 
(\ref{eq:cev_y}) as
\begin{equation}
 n(1+\epsilon) \delta v_y +
  \frac{\epsilon\Omega}{2}\delta v_x=0.
\label{eq:eom_dust_y_2}
\end{equation}
Thus, we can rewrite Equation (\ref{eq:eom_dust_x_l}) as follows,
\begin{equation}
 \frac{\epsilon}{1+\epsilon}\frac{\Omega^2}{n}\delta v_x 
  = 2\pi G \frac{k\epsilon \Sigma_0}{n+Dk^2}\delta v_x-\frac{\delta
  v_x}{\ts}.
\label{eq:eom_dust_x_2}
\end{equation}
(We can derive the same equation from the equation for the centroid
velocity in the x-direction instead of Equation (\ref{eq:eom_dust_x_l}).)
Thus, we obtain the dispersion relation,
\begin{eqnarray}
&& n^2 + \left\{\frac{\epsilon\ts}{1+\epsilon}[\Omega^2 - 2\pi G
	(1+\epsilon\Sigma_0k)] +Dk^2\right\}n\nonumber\\
&& +\frac{\epsilon\ts}{1+\epsilon}\Omega^2 Dk^2 =0.
\end{eqnarray}
This is approximately same as Equation (\ref{eq:desp_2fluid_SGI}) in the
case $\cs^2k^2 \gg \Omega^2 -2\pi G \Sigma_0k$.
The condition for the instability is given by 
\begin{equation}
 \left\{\frac{\epsilon\ts}{1+\epsilon}[\Omega^2 - 2\pi G
	(1+\epsilon\Sigma_0k)] +Dk^2\right\}n <0.
\end{equation}
Therefore, the long wavelength perturbations are stabilized by the term
proportional to $\Omega^2$.
This term comes from the Coriolis force acting on the dust.
Thus we conclude that Coriolis force stabilizes the long wavelength
perturbations.
The relations of the velocities of the gas and the dust 
$\delta v_x \gg \delta u_x$ 
and $\delta v_y \sim \delta u_y$ are essential in 
understanding the effect of the back reaction.
As mentioned above, the radial motion of the gas is suppressed by the
pressure (Equation (\ref{eq:rev_x})). On the other hand, there is no
force to suppress the rotational velocity of the gas. Therefore, the
rotational velocity of the gas is similar to that of the dust (Equation
(\ref{eq:duy_dvy})). The Coriolis force due to the radial velocity of
the dust accelerates the dust in the y-direction
and also gas through friction. Although the left hand side of
Equation (\ref{eq:eom_dust_x_2}) is suppressed by a factor
 $\epsilon/(1+\epsilon)$, this term remains and stabilize the
 long wavelength perturbations.

If the back reaction is neglected, 
the equation of motion of the dust in the y-direction is approximately
given by
\begin{equation}
 \frac{\Omega}{2}\delta v_x = -\frac{\delta v_y}{\ts},
\end{equation}
where we use the terminal velocity approximation ($|n| \ll \ts^{-1}$).
This equation means that the Coriolis force balances the frictional
drag force.
In this case, the long-wavelength perturbations appear to be
 always unstable.
Thus, in \cite{2011ApJ...731...99Y} the maximum
 $\alpha$ for the secular GI is given by the 
condition $n^{-1} =  t_{\rm dri}$.
However, $t_{\rm dri}$ is usually much larger than the dynamical
 timescale. Therefore the back reaction should not be neglected and
it lowers the maximum $\alpha$ for the instability.
\cite{1981A&A....98..173C} also solve the equations for both the gas and
 the dust but they focused on dust rich disks.  
In general, the existence of the growth rate much smaller 
than the dynamical timescale corresponds to the near cancellation 
of various acceleration terms in equation of motion for the unstable
mode.  In this case, even a very small term possibly contributes
significantly to the growth rate, and thus, should not be neglected. 

\section{SUMMARY}
\label{conclusion}
We investigate instability due to gas-dust friction 
in protoplanetary disks by a local analysis
using two-fluid equations for gas and dust.
This instability is reduced to the secular GI 
if we ignore dynamical feedback from dust grains in the gas
equation of motion.
We obtain the approximate dispersion relation, Equation 
(\ref{eq:desp_2fluid_SGI}), and the condition for the instability,
Equation (\ref{eq:app_condition_for_instabiity3}).
We found that the long-wavelength perturbations are stabilized due to
the feedback.
The condition
for the instability of the full treatment of gas and dust is more
difficult to be satisfied than the condition for the secular
GI derived by \cite{2011ApJ...731...99Y} because
of the back reaction of the frictional force from dust to gas.
 In the case $Q\sim 3$, corresponding to a marginally gravitationally 
stable disk, a high dust-to-gas mass ratio,$\epsilon \gtrsim 0.1$, and
week turbulence $\alpha \lesssim 4 \times 10^{-4}$ are required for the 
instability and the growth timescale is about $2\times 10^4$ yr if
 $\epsilon \sim 0.1$.
Since the instability grows 
when dust to gas ratio is enhanced and turbulence is week,
the observation of ring-like structures may be an indicator of a 
dust-concentrated, weekly turbulent disk.

We thank Takashi Nakamura for his continuous encouragement and Hiroshi
Kobayashi, Satoshi Okuzumi and Kohei Inayoshi for fruitful discussion.
We also thank Jennifer M. Stone for useful comments.
This work was supported by Grant-in-Aid for JSPS Fellows.


\begin{thebibliography}{}
\expandafter\ifx\csname natexlab\endcsname\relax\def\natexlab#1{#1}\fi

\bibitem[{{Andrews} {et~al.}(2011){Andrews}, {Wilner}, {Espaillat}, {Hughes},
  {Dullemond}, {McClure}, {Qi}, \& {Brown}}]{2011ApJ...732...42A}
{Andrews}, S.~M., {Wilner}, D.~J., {Espaillat}, C., {et~al.} 2011, \apj, 732,
  42

\bibitem[{{Casassus} {et~al.}(2013){Casassus}, {van der Plas}, {M}, {Dent},
  {Fomalont}, {Hagelberg}, {Hales}, {Jord{\'a}n}, {Mawet}, {M{\'e}nard},
  {Wootten}, {Wilner}, {Hughes}, {Schreiber}, {Girard}, {Ercolano}, {Canovas},
  {Rom{\'a}n}, \& {Salinas}}]{2013Natur.493..191C}
{Casassus}, S., {van der Plas}, G., {M}, S.~P., {et~al.} 2013, \nat, 493, 191

\bibitem[{{Chiang} \& {Youdin}(2010)}]{2010AREPS..38..493C}
{Chiang}, E., \& {Youdin}, A.~N. 2010, Annual Review of Earth and Planetary
  Sciences, 38, 493

\bibitem[{{Coradini} {et~al.}(1981){Coradini}, {Magni}, \&
  {Federico}}]{1981A&A....98..173C}
{Coradini}, A., {Magni}, G., \& {Federico}, C. 1981, \aap, 98, 173

\bibitem[{{Cuzzi} {et~al.}(1993){Cuzzi}, {Dobrovolskis}, \&
  {Champney}}]{1993Icar..106..102C}
{Cuzzi}, J.~N., {Dobrovolskis}, A.~R., \& {Champney}, J.~M. 1993, \icarus, 106,
  102

\bibitem[{{Fukagawa} {et~al.}(2006){Fukagawa}, {Tamura}, {Itoh}, {Kudo},
  {Imaeda}, {Oasa}, {Hayashi}, \& {Hayashi}}]{2006ApJ...636L.153F}
{Fukagawa}, M., {Tamura}, M., {Itoh}, Y., {et~al.} 2006, \apjl, 636, L153

\bibitem[{{Fukagawa} {et~al.}(2013){Fukagawa}, {Tsukagoshi}, {Momose}, {Saigo},
  {Ohashi}, {Kitamura}, {Inutsuka}, {Muto}, {Nomura}, {Takeuchi}, {Kobayashi},
  {Hanawa}, {Akiyama}, {Honda}, {Fujiwara}, {Kataoka}, {Takahashi}, \&
  {Shibai}}]{2013arXiv1309.7400F}
{Fukagawa}, M., {Tsukagoshi}, T., {Momose}, M., {et~al.} 2013, ArXiv e-prints,
  arXiv:1309.7400

\bibitem[{{Geers} {et~al.}(2007){Geers}, {Pontoppidan}, {van Dishoeck},
  {Dullemond}, {Augereau}, {Mer{\'{\i}}n}, {Oliveira}, \&
  {Pel}}]{2007A&A...469L..35G}
{Geers}, V.~C., {Pontoppidan}, K.~M., {van Dishoeck}, E.~F., {et~al.} 2007,
  \aap, 469, L35

\bibitem[{{Goldreich} \& {Lynden-Bell}(1965)}]{1965MNRAS.130..125G}
{Goldreich}, P., \& {Lynden-Bell}, D. 1965, \mnras, 130, 125

\bibitem[{{Hashimoto} {et~al.}(2011){Hashimoto}, {Tamura}, {Muto}, {Kudo},
  {Fukagawa}, {Fukue}, {Goto}, {Grady}, {Henning}, {Hodapp}, {Honda},
  {Inutsuka}, {Kokubo}, {Knapp}, {McElwain}, {Momose}, {Ohashi}, {Okamoto},
  {Takami}, {Turner}, {Wisniewski}, {Janson}, {Abe}, {Brandner}, {Carson},
  {Egner}, {Feldt}, {Golota}, {Guyon}, {Hayano}, {Hayashi}, {Hayashi}, {Ishii},
  {Kandori}, {Kusakabe}, {Matsuo}, {Mayama}, {Miyama}, {Morino}, {Moro-Martin},
  {Nishimura}, {Pyo}, {Suto}, {Suzuki}, {Takato}, {Terada}, {Thalmann},
  {Tomono}, {Watanabe}, {Yamada}, {Takami}, \& {Usuda}}]{2011ApJ...729L..17H}
{Hashimoto}, J., {Tamura}, M., {Muto}, T., {et~al.} 2011, \apjl, 729, L17

\bibitem[{{Hashimoto} {et~al.}(2012){Hashimoto}, {Dong}, {Kudo}, {Honda},
  {McClure}, {Zhu}, {Muto}, {Wisniewski}, {Abe}, {Brandner}, {Brandt},
  {Carson}, {Egner}, {Feldt}, {Fukagawa}, {Goto}, {Grady}, {Guyon}, {Hayano},
  {Hayashi}, {Hayashi}, {Henning}, {Hodapp}, {Ishii}, {Iye}, {Janson},
  {Kandori}, {Knapp}, {Kusakabe}, {Kuzuhara}, {Kwon}, {Matsuo}, {Mayama},
  {McElwain}, {Miyama}, {Morino}, {Moro-Martin}, {Nishimura}, {Pyo}, {Serabyn},
  {Suenaga}, {Suto}, {Suzuki}, {Takahashi}, {Takami}, {Takato}, {Terada},
  {Thalmann}, {Tomono}, {Turner}, {Watanabe}, {Yamada}, {Takami}, {Usuda}, \&
  {Tamura}}]{2012ApJ...758L..19H}
{Hashimoto}, J., {Dong}, R., {Kudo}, T., {et~al.} 2012, \apjl, 758, L19

\bibitem[{{Hayashi}(1981)}]{1981IAUS...93..113H}
{Hayashi}, C. 1981, in IAU Symposium, Vol.~93, Fundamental Problems in the
  Theory of Stellar Evolution, ed. {D.~Sugimoto, D.~Q.~Lamb, \& D.~N.~Schramm},
  113--126

\bibitem[{{Hayashi} {et~al.}(1985){Hayashi}, {Nakazawa}, \&
  {Nakagawa}}]{1985prpl.conf.1100H}
{Hayashi}, C., {Nakazawa}, K., \& {Nakagawa}, Y. 1985, in Protostars and
  planets II (A86-12626 03-90). Tucson, AZ, University of Arizona Press, 1985,
  p. 1100-1153., ed. {D.~C.~Black \& M.~S.~Matthews}, 1100--1153

\bibitem[{{Isella} {et~al.}(2010){Isella}, {Natta}, {Wilner}, {Carpenter}, \&
  {Testi}}]{2010ApJ...725.1735I}
{Isella}, A., {Natta}, A., {Wilner}, D., {Carpenter}, J.~M., \& {Testi}, L.
  2010, \apj, 725, 1735

\bibitem[{{Isella} {et~al.}(2012){Isella}, {P{\'e}rez}, \&
  {Carpenter}}]{2012ApJ...747..136I}
{Isella}, A., {P{\'e}rez}, L.~M., \& {Carpenter}, J.~M. 2012, \apj, 747, 136

\bibitem[{{Isella} {et~al.}(2013){Isella}, {P{\'e}rez}, {Carpenter}, {Ricci},
  {Andrews}, \& {Rosenfeld}}]{2013ApJ...775...30I}
{Isella}, A., {P{\'e}rez}, L.~M., {Carpenter}, J.~M., {et~al.} 2013, \apj, 775,
  30

\bibitem[{{Johansen} \& {Youdin}(2007)}]{2007ApJ...662..627J}
{Johansen}, A., \& {Youdin}, A. 2007, \apj, 662, 627

\bibitem[{{Kley} \& {Nelson}(2012)}]{2012ARA&A..50..211K}
{Kley}, W., \& {Nelson}, R.~P. 2012, \araa, 50, 211

\bibitem[{{Lin} \& {Papaloizou}(1986)}]{1986ApJ...309..846L}
{Lin}, D.~N.~C., \& {Papaloizou}, J. 1986, \apj, 309, 846

\bibitem[{{Lin} \& {Papaloizou}(1993)}]{1993prpl.conf..749L}
{Lin}, D.~N.~C., \& {Papaloizou}, J.~C.~B. 1993, in Protostars and Planets III,
  ed. E.~H. {Levy} \& J.~I. {Lunine}, 749--835

\bibitem[{{Lyra} \& {Kuchner}(2013)}]{2013Natur.499..184L}
{Lyra}, W., \& {Kuchner}, M. 2013, \nat, 499, 184

\bibitem[{{Lyra} \& {Kuchner}(2012)}]{2012arXiv1204.6322L}
{Lyra}, W., \& {Kuchner}, M.~J. 2012, ArXiv e-prints, arXiv:1204.6322

\bibitem[{{Mathews} {et~al.}(2012){Mathews}, {Williams}, \&
  {M{\'e}nard}}]{2012ApJ...753...59M}
{Mathews}, G.~S., {Williams}, J.~P., \& {M{\'e}nard}, F. 2012, \apj, 753, 59

\bibitem[{{Mayama} {et~al.}(2012){Mayama}, {Hashimoto}, {Muto}, {Tsukagoshi},
  {Kusakabe}, {Kuzuhara}, {Takahashi}, {Kudo}, {Dong}, {Fukagawa}, {Takami},
  {Momose}, {Wisniewski}, {Follette}, {Abe}, {Akiyama}, {Brandner}, {Brandt},
  {Carson}, {Egner}, {Feldt}, {Goto}, {Grady}, {Guyon}, {Hayano}, {Hayashi},
  {Hayashi}, {Henning}, {Hodapp}, {Ishii}, {Iye}, {Janson}, {Kandori}, {Kwon},
  {Knapp}, {Matsuo}, {McElwain}, {Miyama}, {Morino}, {Moro-Martin},
  {Nishimura}, {Pyo}, {Serabyn}, {Suto}, {Suzuki}, {Takato}, {Terada},
  {Thalmann}, {Tomono}, {Turner}, {Watanabe}, {Yamada}, {Takami}, {Usuda}, \&
  {Tamura}}]{2012ApJ...760L..26M}
{Mayama}, S., {Hashimoto}, J., {Muto}, T., {et~al.} 2012, \apjl, 760, L26

\bibitem[{{Michikoshi} {et~al.}(2012){Michikoshi}, {Kokubo}, \&
  {Inutsuka}}]{2012ApJ...746...35M}
{Michikoshi}, S., {Kokubo}, E., \& {Inutsuka}, S.-i. 2012, \apj, 746, 35

\bibitem[{{Nakagawa} {et~al.}(1986){Nakagawa}, {Sekiya}, \&
  {Hayashi}}]{1986Icar...67..375N}
{Nakagawa}, Y., {Sekiya}, M., \& {Hayashi}, C. 1986, \icarus, 67, 375

\bibitem[{{Narayan} {et~al.}(1987){Narayan}, {Goldreich}, \&
  {Goodman}}]{1987MNRAS.228....1N}
{Narayan}, R., {Goldreich}, P., \& {Goodman}, J. 1987, \mnras, 228, 1

\bibitem[{{Okuzumi} {et~al.}(2012){Okuzumi}, {Tanaka}, {Kobayashi}, \&
  {Wada}}]{2012ApJ...752..106O}
{Okuzumi}, S., {Tanaka}, H., {Kobayashi}, H., \& {Wada}, K. 2012, \apj, 752,
  106

\bibitem[{{Schmit} \& {Tscharnuter}(1995)}]{1995Icar..115..304S}
{Schmit}, U., \& {Tscharnuter}, W.~M. 1995, \icarus, 115, 304

\bibitem[{{Shakura} \& {Sunyaev}(1973)}]{1973A&A....24..337S}
{Shakura}, N.~I., \& {Sunyaev}, R.~A. 1973, \aap, 24, 337

\bibitem[{{Shu}(1984)}]{1984prin.conf..513S}
{Shu}, F.~H. 1984, in IAU Colloq. 75: Planetary Rings, ed. R.~{Greenberg} \&
  A.~{Brahic}, 513--561

\bibitem[{{Takeuchi} \& {Lin}(2005)}]{2005ApJ...623..482T}
{Takeuchi}, T., \& {Lin}, D.~N.~C. 2005, \apj, 623, 482

\bibitem[{{Takeuchi} {et~al.}(1996){Takeuchi}, {Miyama}, \&
  {Lin}}]{1996ApJ...460..832T}
{Takeuchi}, T., {Miyama}, S.~M., \& {Lin}, D.~N.~C. 1996, \apj, 460, 832

\bibitem[{{van der Marel} {et~al.}(2013){van der Marel}, {van Dishoeck},
  {Bruderer}, {Birnstiel}, {Pinilla}, {Dullemond}, {van Kempen}, {Schmalzl},
  {Brown}, {Herczeg}, {Mathews}, \& {Geers}}]{2013Sci...340.1199V}
{van der Marel}, N., {van Dishoeck}, E.~F., {Bruderer}, S., {et~al.} 2013,
  Science, 340, 1199

\bibitem[{{Vandervoort}(1970)}]{1970ApJ...161...87V}
{Vandervoort}, P.~O. 1970, \apj, 161, 87

\bibitem[{{Ward}(1976)}]{1976fras.conf....1W}
{Ward}, W.~R. 1976, in Frontiers of Astrophysics, ed. E.~H. {Avrett}, 1--40

\bibitem[{{Ward}(2000)}]{2000orem.book...75W}
{Ward}, W.~R. 2000, {On Planetesimal Formation: The Role of Collective Particle
  Behavior}, ed. R.~M. {Canup}, K.~{Righter}, \& {et al.}, 75--84

\bibitem[{{Youdin} \& {Johansen}(2007)}]{2007ApJ...662..613Y}
{Youdin}, A., \& {Johansen}, A. 2007, \apj, 662, 613

\bibitem[{{Youdin}(2011)}]{2011ApJ...731...99Y}
{Youdin}, A.~N. 2011, \apj, 731, 99

\bibitem[{{Youdin} \& {Goodman}(2005)}]{2005ApJ...620..459Y}
{Youdin}, A.~N., \& {Goodman}, J. 2005, \apj, 620, 459

\bibitem[{{Youdin} \& {Lithwick}(2007)}]{2007Icar..192..588Y}
{Youdin}, A.~N., \& {Lithwick}, Y. 2007, \icarus, 192, 588

\bibitem[{{Zhu} {et~al.}(2011){Zhu}, {Nelson}, {Hartmann}, {Espaillat}, \&
  {Calvet}}]{2011ApJ...729...47Z}
{Zhu}, Z., {Nelson}, R.~P., {Hartmann}, L., {Espaillat}, C., \& {Calvet}, N.
  2011, \apj, 729, 47

\end{thebibliography}
\end{document}